\documentclass[10pt,conference]{IEEEtran}
\IEEEoverridecommandlockouts

\usepackage{cite}
\usepackage{amsmath,amssymb,amsfonts}
\usepackage{graphicx}
\usepackage{textcomp}
\usepackage{xcolor}
\usepackage{tabularx}

\usepackage{hyperref}
\usepackage{url}
\usepackage{algpseudocode}
\usepackage{algorithm}
\usepackage{tabularx,booktabs}
\usepackage{graphicx}
\usepackage{subcaption}

\usepackage{tablefootnote}
\usepackage{tikz}
\usepackage{float}
\usepackage{circledsteps}
\newcommand{\circled}[1]{\tikz[baseline=(char.base)]{\node[shape=circle,draw,inner sep=1.5pt](char){#1}}}

\def\BibTeX{{\rm B\kern-.05em{\sc i\kern-.025em b}\kern-.08em
    T\kern-.1667em\lower.7ex\hbox{E}\kern-.125emX}}
\begin{document}

\title{GreenWhisk: Emission-Aware Computing for Serverless Platform}

\author{
\IEEEauthorblockN{Jayden Serenari\textsuperscript{*}\textsuperscript{1}, Sreekanth Sreekumar\textsuperscript{*}\textsuperscript{1}, Kaiwen Zhao\textsuperscript{*}\textsuperscript{1}, Saurabh Sarkar\textsuperscript{2}, Stephen Lee\textsuperscript{1}}
\IEEEauthorblockA{\textsuperscript{1}\textit{University of Pittsburgh, \textsuperscript{2}TryCarbonara}}
\thanks{\textsuperscript{*}These authors contributed equally to this work.}

}

\maketitle

\begin{abstract}
Serverless computing is an emerging cloud computing abstraction wherein the cloud platform transparently manages all resources, including explicitly provisioning resources and geographical load balancing when the demand for service spikes. Users provide code as functions, and the cloud platform runs these functions handling all aspects of function execution. While prior work has primarily focused on optimizing performance, this paper focuses on reducing the carbon footprint of these systems making variations in grid carbon intensity and intermittency from renewables transparent to the user. We introduce GreenWhisk, a carbon-aware serverless computing platform built upon Apache OpenWhisk, operating in two modes — grid-connected and grid-isolated — addressing intermittency challenges arising from renewables and the grid's carbon footprint. Moreover, we develop carbon-aware load balancing algorithms that leverage energy and carbon information to reduce the carbon footprint. Our evaluation results show that GreenWhisk can easily incorporate carbon-aware algorithms, thereby reducing the carbon footprint of functions without significantly impacting the performance of function execution. In doing so, our system design enables the integration of new carbon-aware strategies into a serverless computing platform. 

\end{abstract}


\section{Introduction}

The growth of cloud platforms has resulted in an increased demand for energy as they strive to meet the requirements driven by data-intensive workloads, such as machine learning~\cite{strubell2020energy, patterson2021carbon}. Recent studies indicate that the energy needs of cloud data centers currently account for 2\% of global carbon emissions~\cite{masanet2020recalibrating} and may increase to 8\% by 2030~\cite{andrae2015global}. This is attributed to the fact that, over the past decade, enhancements in energy efficiency have kept pace with capacity growth, resulting in a modest increase in energy consumption. However, opportunities for further energy efficiency within computing systems are now limited, raising concerns about the potential for escalating energy usage and associated carbon emissions~\cite{souza2023ecovisor, gupta2021chasing}. 

To reduce carbon emissions, a key design imperative is to optimize for carbon efficiency, i.e., carbon emissions per work done. Achieving carbon efficiency goes beyond energy efficiency, as even data centers optimized for energy efficiency can have substantial carbon emissions if powered by non-renewable sources. Acknowledging this, large technology companies increasingly prioritize clean energy sources and invest in renewable energy to power their data centers~\cite{xing2023carbon} sustainably. 

While significant efforts have been dedicated to optimizing energy and carbon efficiency at the system  or  application level~\cite{sharma2023challenges, james2019low, souza2023ecovisor, xing2023carbon, shen2013power,rezaei2021energy}, there has been little work in addressing the carbon efficiency of existing cloud platforms. This gap has motivated us to reconsider the design of cloud platforms with a focus on operating in a carbon-free environment.  In this paper, we focus on decarbonizing serverless computing platforms by introducing new carbon-aware abstractions geared toward enhancing the carbon efficiency of applications. We have chosen serverless computing as it represents an emerging cloud computing model wherein users do not have to manage cloud resources (e.g., containers). Instead, users define functions, and the cloud provider manages all aspects of function execution, including resource provisioning. This shift in responsibility may increase the overall energy consumption of serverless functions compared to traditional self-managed services. Prior research indicates that introducing additional control mechanisms to manage function execution can result in a nearly 15x increase in consumption compared to traditional self-managed services~\cite{sharma2023challenges}.

The main challenge in designing carbon-aware abstractions is that most interface designs provide controls at the system level (e.g., containers), leaving applications to determine how to handle scenarios efficiently during high carbon periods or when no energy is available. This setup leaves applications responsible for determining how to handle scenarios efficiently during high carbon periods or when no energy is available.  Addressing this challenge involves not only setting emission targets but also understanding the variations in energy availability and carbon intensity. This multifaceted approach is crucial for effectively managing carbon in serverless computing platforms' dynamic and diverse landscape.

We note that awareness of carbon allows serverless computing platforms to optimize function invocation scheduling to reduce emissions. Furthermore, understanding grid carbon and energy allows easy integration of different carbon-management policies for applications to reduce emissions. We redesigned Apache OpenWhisk~\cite{openwhisk}, a popular open-source platform used by several companies, to validate our hypothesis and implement carbon-aware abstractions that handle energy and carbon intensity intermittency. We are not aware of any prior work on rearchitecting serverless platforms as a whole to reduce emissions.  In doing so, we make the following contributions.

\begin{itemize}
  \item GreenWhisk Design: We redesign OpenWhisk to incorporate mechanisms that enable applications to optimize carbon-efficiency.
We also introduce energy interface API to expose the underlying energy profile to load balancing algorithms. This enables the implementation of new carbon-aware load algorithms that optimize carbon efficiency. 
The design enables Green Whisk to run on grid-connected and grid-isolated modes, making the underlying carbon intermittency transparent to users. 
 
\item Carbon-aware Load Balancer: We also implement a carbon-aware load balancer that is both locality and carbon-aware. This helps in optimizing the carbon-efficiency.  We also introduce mechanisms that allow invocation requests to be scheduled later when renewable energy is available.

    \item Implementation and Evaluation: We evaluate the performance of the GreenWhisk prototype on both a server and Raspberry Pi cluster. Furthermore, we also simulate the performance and emission savings for a one-year period. Our results show that GreenWhisk reduces the downtime by 50\% compared with the default OpenWhisk and avoids twice as many emissions as the baseline. The code will be available on Github.
\end{itemize}

\section{Background}

\noindent
{\bf Serverless Platform Architecture.}
The serverless platform, referred to as Function as a Service (FaaS), provides users with transparency in resource management. In this model, users register their application code as functions on a FaaS platform and invoke them through service endpoints. These service endpoints, managed by the FaaS platform, efficiently route function execution requests to servers responsible for hosting the respective functions. A key characteristic of serverless functions is their automatic scaling in direct response to the number of incoming requests. The FaaS platform handles the complexities of executing these functions, providing authentication, function isolation, and dynamic resource management.

\noindent
{\bf OpenWhisk Function Workflow.}
The OpenWhisk platform follows a similar architecture, where users register functions as Actions via an API gateway. Key components, including the \textit{controller}, \textit{invoker}, \textit{Apache Kafka}, and the \textit{Database (CouchDB)}, collaboratively orchestrate the serverless computing environment. The \textit{controller} plays a key role in processing user requests, managing authentication and authorization, and directing requests to appropriate resources. The controller also coordinates serverless function execution and maintains the platform state. \textit{Invokers} execute individual actions (i.e., user functions), dynamically scaling to meet demand. Each invoker manages multiple ephemeral Docker containers, ensuring isolation and supporting multitenancy. \textit{Apache Kafka} acts as a message broker, facilitating communication between invokers and the controller, ensuring efficient orchestration of the serverless computing environment. Finally, \textit{CouchDB}, a NoSQL database, stores metadata encompassing actions, code, results, and authentication details. 
\begin{figure}[t]
    \centering
    \includegraphics[scale=0.4]{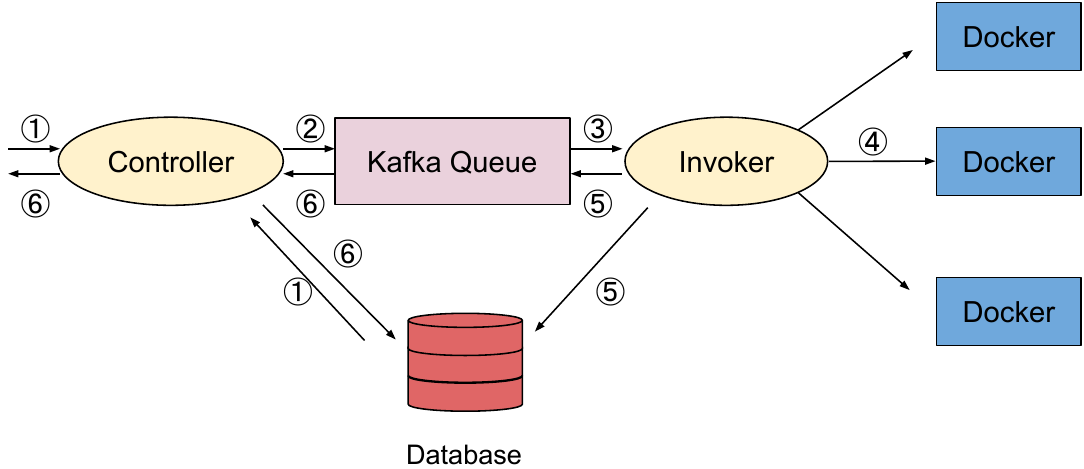}
    \caption{OpenWhisk workflow.}
    \label{fig:openwhisk}
\end{figure}

The OpenWhisk function workflow, depicted in Figure~\ref{fig:openwhisk}, begins with the user invoking a function through a REST API call to the controller. \circled{1}\ The controller, in turn, authenticates the user and retrieves the necessary source code for execution from the database. The controller returns an activation ID for asynchronous invocation, allowing the user to access the result later. \circled{2} Using a load balancer algorithm, the controller selects an invoker and dispatches the invocation request to the invoker's Kafka topic. \circled{3} Upon receiving the invocation request, the invoker selects a Docker container, injects the code, and executes it. \circled{4} The invoker also decides whether to reuse an existing container or create a new one for function invocation.  \circled{5} Subsequently, the invoker communicates the function execution result of the invocation back to the controller, concurrently storing the result in the database. \circled{6} The controller receives the result from Kafka and transmits it to the user in the case of a synchronous invocation.

\noindent
{\bf Load Balancing Locality versus Carbon Intensity.} 
FaaS platforms use a load balancer to distribute function invocation requests among worker nodes. Prior work has proposed various methods for assigning invocation requests --- a common approach being to make them locality aware~\cite{fuerst2022locality}. Here, \textit{locality} refers to executing the function on the same server. By assigning functions to the same worker, this approach mitigates the cold-start problem, as there may already be a container in operation. One common technique for achieving locality is using a consistent hashing algorithm~\cite{aumala2019beyond}. Functions and servers are hashed to points in a circle, and functions are assigned to the server in the next clockwise direction. Functions are sent to the next clockwise server if the server is busy or unavailable.

OpenWhisk utilizes a modified consistent hashing algorithm within its load balancer. Specifically, the controller tracks the memory footprint of each server node using a local counter where functions are executed. When a function is assigned to an invoker, the controller reduces the memory size allocation for that invoker, and it is subsequently increased once the function completes execution. If the memory limit for a specific server is reached, the controller randomly redirects the invocation request to another server.

\begin{figure}[t]
    \centering
    \includegraphics[width=3.1in]{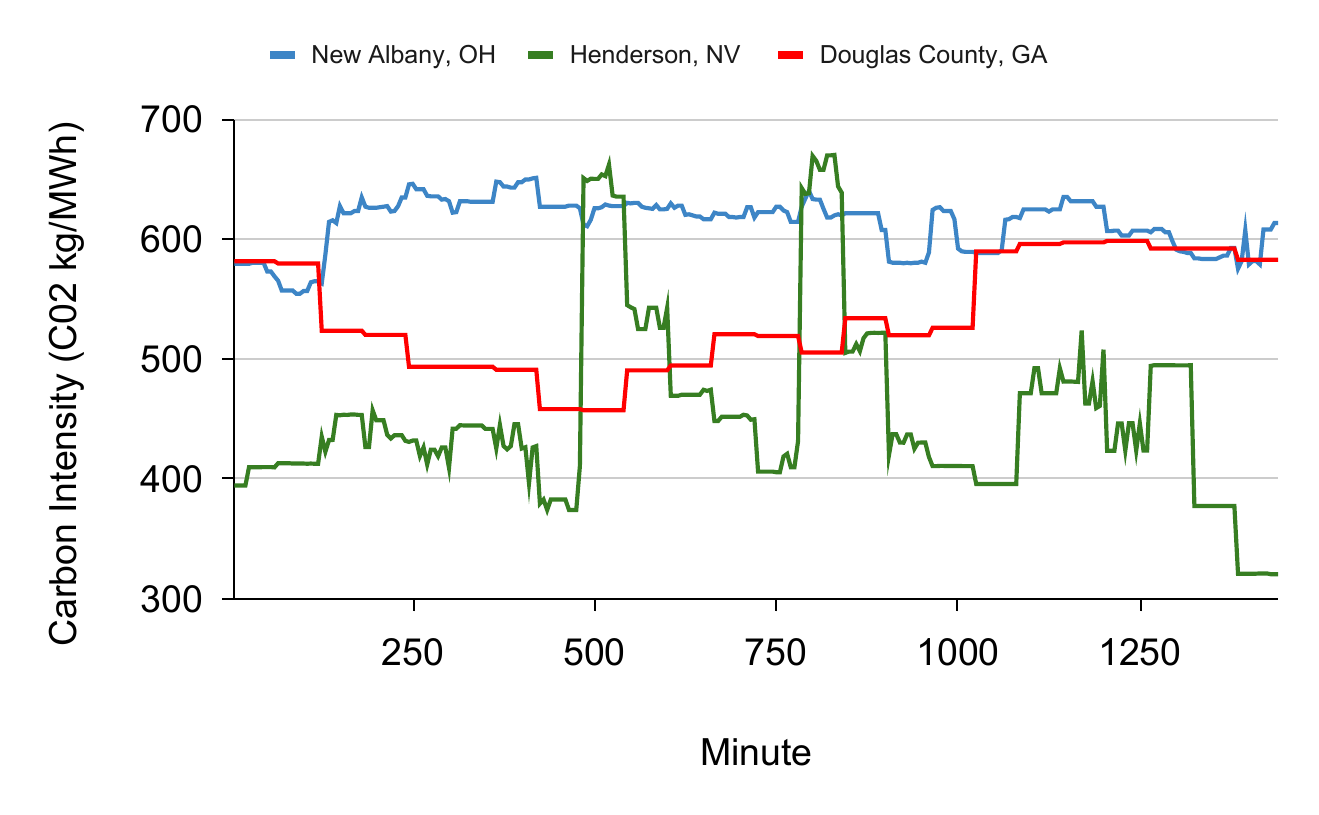}
    \caption{Variations in carbon intensity across different US cities.}
    \label{fig:moer_sample}
\end{figure}

While locality reduces the overhead of function initialization, optimizing for carbon efficiency poses a distinct challenge. 
The carbon footprint is closely tied to the energy sources of the server.

If the server relies on grid energy, its carbon footprint is influenced by the mix of generators used to meet the demand. The grid employs a combination of generators to balance supply and demand in real time dynamically. Depending on the fuel type, grid generators may exhibit a wide range of carbon emissions. 
 
This variability may also differ across different locations.
As depicted in Figure~\ref{fig:moer_sample}, grid emissions demonstrate fluctuations across diverse locations and time periods. 
The load balancer needs to factor in these variations to optimize for carbon efficiency. However, exclusively optimizing for carbon efficiency may result in numerous cold starts, incurring high overall function execution latency. Thus, a key challenge lies in balancing locality and carbon efficiency.

Managing invocation requests on a FaaS platform that relies on intermittent renewable energy, such as solar power, presents additional challenges. While the carbon footprint from renewable energy is zero, the energy supply's intermittent nature can lead to servers' temporary unavailability. A key challenge in designing a load balancer is optimizing function execution despite disparities in energy availability across servers. Importantly, power generation and energy consumption constraints are independent. Power generation depends on environmental factors, such as sunlight for solar power, while the workload may vary depending on demand and application requirements. Most serverless platforms, including OpenWhisk, assume that servers are always on and not designed to operate on intermittent energy, creating a potential mismatch with the intermittent energy supply model.

\noindent
{\bf Assumptions.}
Our work assumes that the serverless platform operates in either \textit{grid-connected} or \textit{grid-isolated} mode. In grid-connected mode, servers are powered by grid energy, whereas in the grid-isolated model, servers are entirely disconnected from the grid, relying on on-site renewables and utilizing batteries to store energy. While grid-connected data centers can incorporate local renewable energy sources, we make this distinction because the grid-isolated mode introduces additional challenges due to potential energy limitations. In the case of a serverless platform operating on both the grid and solar power, we can combine the carbon intensity of both sources to calculate the overall emission footprint.

Our work also assumes we can monitor the server's energy and carbon footprint in real time. Existing server components already expose their energy consumption information to the operating system, such as IPMI~\cite{ipmioverview}. Additionally, battery charge controllers and solar inverters can provide details on energy levels and solar power generation.

We can also monitor the carbon footprint of the energy sources in real time. Third-party web services offer real-time carbon intensity data for various locations, such as WattTime~\cite{watttime} and Electricity Map~\cite{elecmap}. Existing studies are now using these services to estimate their carbon footprint. We base our technique on these assumptions and presume that energy and carbon emissions are accessible.

\section{GreenWhisk Design}

\begin{figure*}[th]
    \centering
    \includegraphics[width=6.2in]{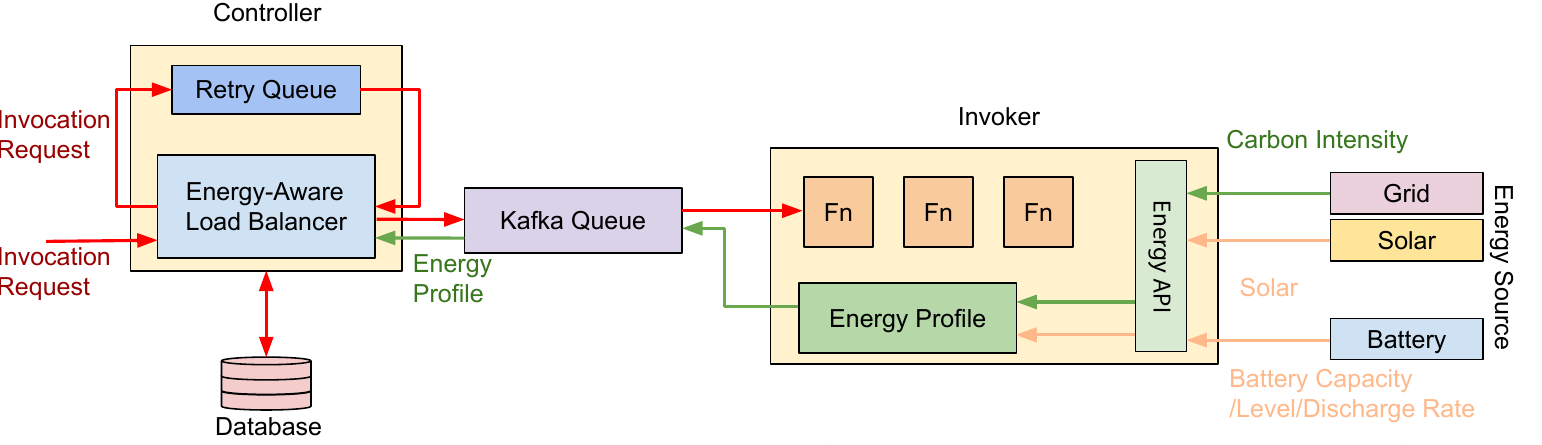}
    \caption{GreenWhisk system architecture. The system can operate on either grid-connected or grid-isolated mode.}
    \label{fig:greenwhiskdesign}
\end{figure*}
\label{design}
This section describes the system architecture of GreenWhisk
and the carbon-aware load balancer.  

\subsection{System Architecture}
Our goal in designing GreenWhisk was to make the underlying carbon intermittency transparent to the user and enable FaaS platform to optimize carbon efficiency under various scenarios. We opted to support both grid-connected and grid-isolated modes, each presenting unique challenges in terms of carbon- and energy-intermittency, respectively. Figure~\ref{fig:greenwhiskdesign} provides an overview of the overall system design, highlighting our extensions to GreenWhisk. We discuss the details in the following subsections.

\begin{table}[t]
  \centering
  \small
  \caption{GreenWhisk's Energy Interface API for grid-connected and grid-isolated modes.}
  \label{tbl:gridconn}
  \begin{tabularx}{\columnwidth}{l|l}     \toprule
    \textbf{Function Name} &  \textbf{Description} \\    \midrule
    \texttt{get\_carbon\_intensity()}  & Get grid carbon intensity\\\hline
    \texttt{get\_battery\_level()} & Get available battery 
     energy (\%) \\\hline
    \texttt{get\_battery\_capacity()} & Get overall energy capacity \\\hline
    \texttt{get\_discharge\_rate()} & Get battery's max 
    discharge rate \\\hline
    \texttt{get\_solar()} & Get solar energy \\\bottomrule

  \end{tabularx}
\end{table}

\subsubsection{Energy Interface}
\label{API}
Table \ref{tbl:gridconn} outlines the GreenWhisk energy interface for grid and grid-isolated modes. When connected to the grid, GreenWhisk utilizes these methods to gather the grid's carbon intensity. As discussed, this real-time information is easily accessible through existing APIs, which the interface leverages. In grid-isolated mode, GreenWhisk uses the API to retrieve information on solar power output, battery energy availability, battery capacity, and the battery discharge rate. This data is crucial for understanding the energy available to the servers to service the current and future workload.

\subsubsection{Energy Profile}
This component operates on each server, monitoring the carbon footprint and available energy. At specific intervals, the Energy Profile retrieves energy and carbon intensity data, creates a timestamped topic, and sends it to the Controller via the Kafka queue. The default interval is set to 1 second. We use Kafka to communicate this message to the controller, and the timestamped message is needed to understand the energy and carbon footprint across various servers at any given time. We assume that servers synchronize their clocks using protocols such as NTP.

However, the impact of clock synchronization or the delay in receiving messages by the controller is negligible and does not significantly affect function execution. This is analogous to the load balancer receiving slightly outdated information to make decisions. In the grid-connected scenario, energy is consistently available via the grid. Thus, if a function is assigned to an invoker by the load balancer, the invoker will execute it if the invoker is healthy. Nonetheless, this may have implications for overall emissions.  In the grid-isolated scenario, this may become an issue if the delay in receiving messages is substantial (e.g., hours). Essentially, there is a possibility that the controller receives delayed energy information, indicating that energy is available when, in reality, there is none. This situation is akin to when the controller assumes that the invoker is healthy, but the server is down. In such cases, the invoker may fail to execute the function.

To address this issue in the grid-isolated mode, GreenWhisk maintains a minimum energy buffer in the battery at each server location. This energy buffer acts as a stopgap, allowing the servers to continue functioning even if there is a delay in receiving messages by the controller. The energy buffer serves as a mechanism to handle the maximum delay (staleness) the system can tolerate. 
Let $P^{max}_s$ denote the power drawn by the server $s$ at maximum capacity. We can compute the server's  operation time as follows:
\begin{equation}
    op\_time_s = \frac{energy\_buffer_s} {P^{max}_s} 
\end{equation}
In our implementation, we configure the energy buffer to be a fraction of the maximum battery capacity, for example, 20\% of the battery capacity. Importantly, $op\_time_s$ represents the maximum delay the system can tolerate. If the controller assigns a function within the server's operation time, there is a high likelihood that the function will be executed before the server exhausts its energy. 

To illustrate, let us consider a scenario with a 5kWh battery and a maximum power draw of 0.5kW. If the energy buffer is set to 20\%, the server can operate for 2 hours within its operation time. Note that the operation time will decrease with a higher number of servers. Despite having an energy buffer, there remains the possibility of an invoker being assigned a function even when there is no energy. 
In such instances, function execution is likely to fail. GreenWhisk gracefully handles this situation by reverting to the default OpenWhisk implementation and utilizing a timeout mechanism to notify users if a function fails. This ensures that users are informed promptly in case of any execution failures.

\subsubsection{Retry Queue}
This component facilitates scheduling any function execution at a later time. 
In grid-isolated mode, where there is no energy across servers, the controller faces challenges in assigning a server for function invocation. A common strategy is to fail the invocation request and promptly notify the user. This approach aligns with the OpenWhisk design, where the platform reports a failure when all servers are down. In contrast, our design incorporates a Retry Queue that enables the controller to schedule the function later when renewable energy is available. The Retry Queue introduces more flexibility in determining when the function can be executed, enabling the load balancing algorithm to leverage temporal variations in carbon and energy availability. 

To provide further control, we incorporate a parameter that dictates how often a function can be placed in the Retry Queue before notifying the user. Users receive immediate notifications of failure when this value is set to zero. However, in our implementation, we set the default value to 3, enabling the load balancer to exploit temporal variations. While we employ the Retry Queue mechanism when energy is unavailable in grid-isolated mode, it can also be leveraged in grid-connected mode to exploit temporal variations in carbon intensity. Specifically, the invocation request may remain in the queue during periods of high carbon intensity and be processed from the queue during low carbon periods. 

\subsubsection{Controller}
The GreenWhisk controller serves as the central orchestrator within the system, managing services such as function invocation requests, load balancing, and resource management. We enhance the controller's capabilities by integrating features that promote energy and carbon awareness. Specifically, we adapted the controller to collect Energy Profile messages from Kafka, thereby providing valuable energy and carbon intensity data to the load balancer. This integration enables the load balancer to make informed decisions based on the current energy and carbon conditions across servers. Furthermore, our modifications extend to exposing the RetryQueue mechanism to the load balancer. This enhancement allows the load balancer to interact with the Retry Queue, enabling it to schedule invocation requests intelligently later.

\begin{algorithm*}
\caption{Carbon-aware Load Balancer}\label{alg:one}
\begin{algorithmic}[1]
\Procedure{distance}{$server$, $function$}
  \State \Return $-\ln(\ ( 1 - (server - function) )\mod 1\ )$
\EndProcedure
\Procedure{weight\_grid\_isolated}{$i$, $avail\_energy$}
  \State \Return $avail\_energy_i / \sum_j avail\_energy_j $
\EndProcedure
\Procedure{weight\_grid\_conn}{ $i$, $carbon\_intensity$ }
  \State \Return $- carbon\_intensity_i / \sum_j carbon\_intensity_j $
\EndProcedure
\Procedure{sort\_servers}{$servers$, $function$, $metric$, $weight\_fun$}
    \State $s\_list \gets [\ (s,\ distance(s, function)\ /\ weight\_fun(s, metric))\ \textbf{for}\ s \in servers] $
    \State $s\_list \gets s\_list.sort(key=operator.itemgetter(1)) $
    \State \Return $s\_list$
\EndProcedure
\Procedure{select\_server}{$servers$, $function$, $metric$}
    \State $c \gets None$
    \For{$s \in sort\_servers(servers, ...)$}
        \If {$s.memory < mem\_limit$}
            \State $c \gets s$
            \State \textbf{break}
        \EndIf
    \EndFor
    \If{$c$ \textbf{is} $None$}
        \State $retry\_queue.add(function)$
    \Else
        \State $s.invoke(function)$
    \EndIf
\EndProcedure
\end{algorithmic}
\end{algorithm*}

\subsection{Carbon-aware Load Balancer}
\label{Algorithms}
Our objective is to minimize overall function execution latency and enhance the carbon efficiency of the system in both grid-connected and grid-isolated modes. However, existing FaaS algorithms mainly focus on locality awareness, which reduces end-to-end function latency. 
As discussed, while consistent hashing achieves locality, it does not consider energy or carbon in routing functions. This motivates our energy and carbon-aware load balancing algorithm, which we describe below.

GreenWhisk's load balancer builds on a variant of weighted consistent hashing to balance high locality and carbon awareness within the system~\cite{schindelhauer2005weighted}. The key idea of weighted consistent hashing involves assigning weights to each server, reflecting the inclination toward assigning functions to servers. Servers with a lower carbon footprint or higher availability of green energy are assigned higher weights. Consequently, servers with more weights are assigned more hash ranges and, thus, a larger portion of the function space.

We present the carbon-aware load balancer in Algorithm~\ref{alg:one}. GreenWhisk's load balancer assigns functions to servers using a two-step process. Initially, the algorithm calculates the distance between a function's hash value and a server's hash value (Line 2). This distance metric ensures locality and that functions are consistently mapped to the nearest server.  Subsequently, the algorithm adjusts the distance value based on a specified metric, such as available carbon-free energy or carbon intensity. In the grid-connected mode (Line 7), the algorithm adjusts the distance according to the carbon intensity metric. This adjustment involves dynamically increasing or decreasing the distance relative to the carbon intensity of the server. The key idea is to amplify the distance for servers and functions located in regions with higher carbon intensity and reduce it for those in areas with lower intensity, contributing to a carbon-aware load-balancing strategy.
In the grid-isolated mode, the distance value is adjusted using the available energy metric. The algorithm calculates the available energy by factoring in solar energy, battery discharge, and the energy footprint of functions executed on the server. 
\begin{align*}
  avail\_energy_s (t) = & \ batt\_discharge_s (t) + solar_s (t) \\
  &  - \sum_f energy_f  (t)
\end{align*}
where $batt\_discharge_s (t)$ is the energy discharged by the battery at time $t$, $solar_s (t)$ is the solar energy generation at server $s$ at time $t$, and $energy_f (t)$ is the energy consumed by functions currently running in the server at time $t$. 
Subtracting the energy used by running functions ensures the algorithm avoids system overloading in scenarios with insufficient energy to handle the workload.

Note that OpenWhisk relies on memory limits to address capacity constraints. This may be insufficient in the grid-isolated case where function execution depends on energy availability. In other words, even with adequate computing resources, the functions will fail to execute if there isn't enough energy to power the servers. Recognizing this, our algorithm incorporates the energy footprint of the function into its decision-making process, ensuring that the load balancer selects a location with sufficient energy resources.
In scenarios where servers lack the necessary resources or energy, the algorithm proceeds to the next available server in a sorted list. If, however, there are no available servers, the algorithm places the function into the Retry Queue. GreenWhisk then employs a configurable interval before attempting to schedule the function again. This mechanism allows the system to handle scenarios where a server may not be immediately available due to a lack of computing resources or energy.

\section{Implementation}

\label{implementation}
We implemented various components of GreenWhisk, including different load balancers such as a carbon-aware load balancer, consistent hashing, and a greedy algorithm which is exposed as a configuration variable. The implementation was written in Scala within the OpenWhisk framework, resulting in approximately 2870 lines of code changes. The Retry queue was implemented using a Redis-based priority queue with priority inversely proportional to the time the function was first added to the queue. To facilitate energy and carbon monitoring, we exposed the API (Table \ref{tbl:gridconn}) to the Energy Interface, enabling the monitoring of underlying energy and carbon data. 

However, we did not have access to solar array emulators for emulation purposes for the grid-isolated scenario. Instead, we developed energy adaptors to interface with solar generation data, effectively simulating solar panels. Additionally, we designed an adaptor to emulate battery behavior, encompassing charge and discharge operations. To ensure accurate energy modeling, we profiled the energy consumption of Raspberry Pi under various loads and utilized the standard linear model to calculate the power consumption~\cite{mathew2012energy}:
\begin{equation}
    \text{{power}} = P_{\text{{idle}}} + \lambda  \cdot (P_{\text{{peak}}} -   P_{\text{{idle}}}) 
\end{equation}
where the load $0 \leq \lambda \leq 1$ represents the server's resource utilization, $P_{\text{{idle}}}$ is the power consumed by an idle server, and  $P_{\text{{peak}}}$ is the power consumed by the server under peak
load. 

We also developed a GreenWhisk simulator in Java to emulate the behavior of various algorithms for longer-running experiments. The simulator is designed to simulate independent servers, assign functions to servers, and model both cold and warm starts while considering the capacity and energy constraints of the servers. Moreover, we use the power model discussed above to simulate the energy consumption of servers. In the simulation, workloads are represented as Java objects, utilizing the built-in \texttt{Object.hashCode()} function for balancing purposes. To differentiate between cold and warm starts, each server in the simulation is equipped with three simulated containers. The function type (representing Node, Python, etc.) is randomly assigned, and the number of function types is configurable. However, in our experiments, we set the number of function types to 5. For repeatability, the simulation uses a fixed random number seed for the experiments presented in this paper.

We determine whether a function is a warm or cold start based on the function type. If a free container with the same function type is available, it is considered a warm start. Otherwise, the container type is changed to match the function type, and the start is considered cold. Since the time granularity of the workload trace and simulation differs, any remainder of the functions in the trace are evenly spread out across the minute. This strategy also aligns with the one used during emulation.

\section{Evaluation}


\begin{table*}
    \centering
    \small
    \caption{Summary of server locations, their carbon intensity, and solar availability.}
    \label{tbl:my_label}
    \begin{tabular}{|l|c|c|c|}\toprule
         \textbf{Location} & \textbf{Avg. MOER} & \textbf{Avg. GTI} & \textbf{Balancing Authority} \\\midrule
         Henderson, Nevada & 991 & 271 & NEVP \\\hline
         The Dalles, Oregon & 1068 & 201 & BPA \\\hline
         Douglas County, Georgia & 1169 & 203 & SOCO \\\hline
         New Albany, Ohio & 1283 & 174 & PJM \\\hline
         Storey County, Nevada & 991 & 265 & NEVP \\\hline
         Montgomery County, Tennessee & 1139 & 192 & TVA \\\hline
         Papillion, Nebraska & 1108 & 211 & SPP (WEST NE) \\\hline
         Midlothian, Texas & 1099 & 216 & ERCOT (North-Central) \\\hline
         Mayes County, Oklahoma & 1350 & 204 & SPP (Memphis) \\\bottomrule
    \end{tabular}
    
\end{table*}

\label{methodology}
\subsection{Hardware Setup}

\begin{figure}[t]
    \centering
    \includegraphics[scale=0.05,trim={0 0 0 {20cm}},clip]{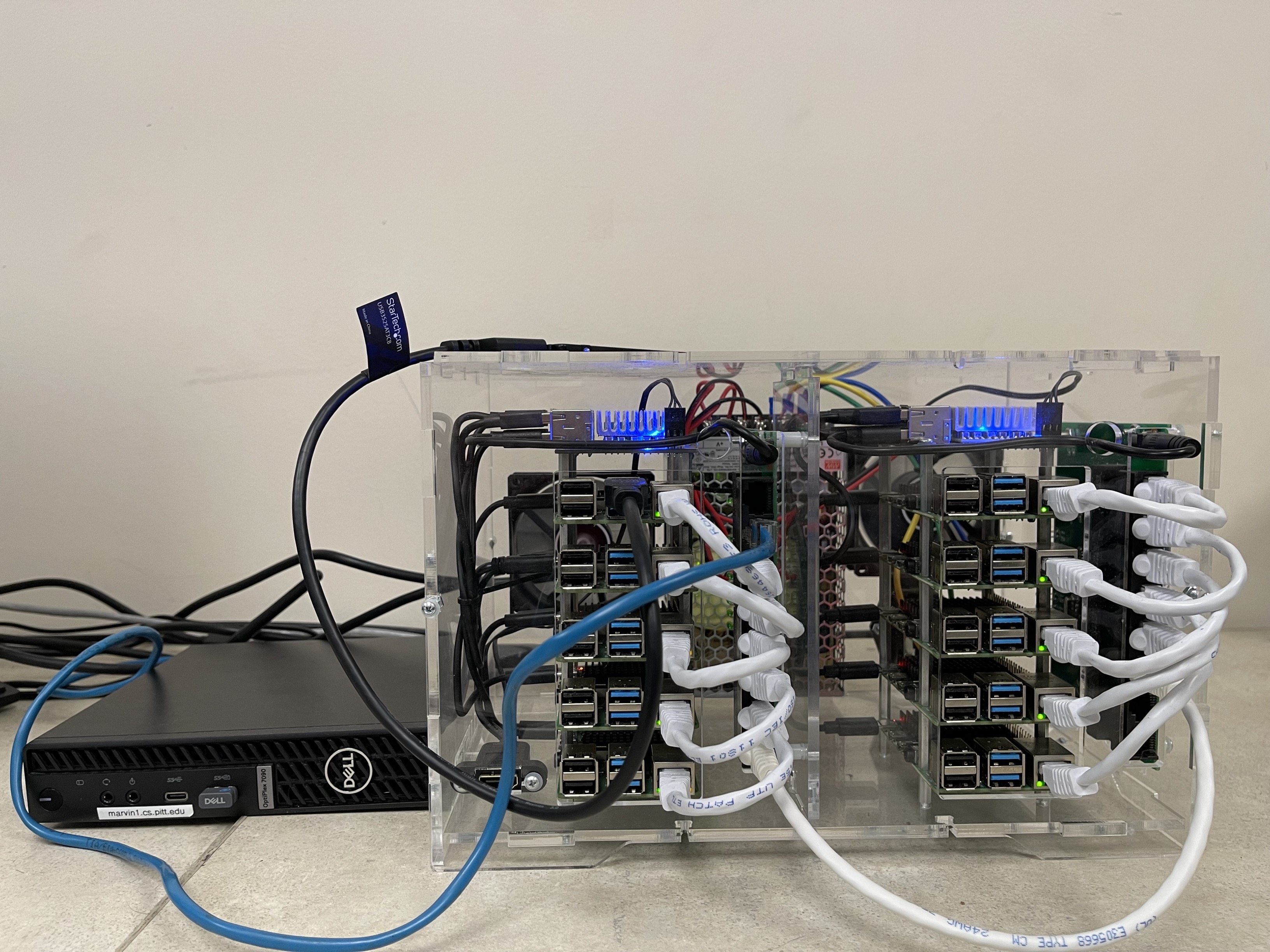}
    \caption{GreenWhisk Raspberry PI FaaS cluster.}
    \label{graph:hardware}
\end{figure}

We conduct our experiments both on a Raspberry Pi cluster and a server cluster. The Pi cluster consists of 10 Raspberry Pi 4 Model B devices with 8GB RAM (Figure \ref{graph:hardware}). The server cluster consists of 10 machines running CentOS, equipped with the 10-core Intel Xeon Silver 4114 processor and 12x32GB DDR4 memory. For our Pi experiments, a desktop machine functions as the controller, orchestrating the deployment and execution of functions on the invokers. The controller machine is an amd64 system with an 8-core 1.4 GHz 11th Gen Intel Core i7 processor and 16GB DDR4 memory. The Raspberry Pi devices serve as invokers where functions are executed. We used Ansible to deploy OpenWhisk on our invokers to facilitate this. We created multi-arch Docker images (armv7 and amd64) for the invoker and function runtimes (Python, Node, and Ruby) and deployed them to the machines through an Ansible build automation.
\subsection{Datasets}

\noindent
{\bf Azure Public Dataset.}
\label{azure} 
The Azure Public Dataset~\cite{shahrad2020serverless} contains data relating to functions and their relative invocations per minute. Because of the memory and CPU inequality between systems, we sampled a subset for our evaluation. Sampling was done by determining the maximum number of function invocations per minute that our system can handle without timing out.

\noindent

\noindent
{\it Rare.}
A random sample of 125 functions that are infrequently invoked with 15 invocation requests per minute.

\noindent
{\it Medium.} A random sample of 50 functions with 54 invocation requests per minute.

\noindent
{\it High.}
A random sample of 50 distinct functions with 354 invocation requests per minute.

\begin{table}
    \centering
    \small
    \caption{Function Trace Statistics.}
    \label{tbl:functrace}
    \begin{tabular}{|l|c|c|c|c|}\toprule
         Function Trace & \#Functions & \#Invoc. & Reqs/min & Avg. IAT \\\midrule
         Rare    &   125    &  451          &      15        &    4.0 s \\\hline
         Medium   &    50     & 1,616        &         54     &   1.1 s    
         \\\hline
         High  &    50     & 10,632        &         354     &   0.2 s
    \\\bottomrule
    \end{tabular}
    
\end{table}

\noindent
{\bf Solcast Solar Data.}
We collected solar irradiance data from the Solcast API~\cite{solcast} to determine our solar profiles. The panel output was estimated and then normalized to a 1-Watt panel. This was then scaled linearly to meet our system specifications. Solar data was provided as global tilted irradiance (GTI) in W/m2. 

\noindent
{\bf Marginal Emissions Dataset.}
We use the WattTime~\cite{watttime} marginal emissions dataset to calculate avoided emissions when switching workloads from one server to another. 
Marginal Operating Emissions Rate (MOER) is a metric used to derive the amount of carbon dioxide  that would be avoided if a load was moved from one power source to another. This unit is given in lbs/MWh. We can scale the MOER to calculate the emissions per function $f$ on server $s$ as follows. 
\begin{equation}
MOER_{\text{scaled}} (s, f) = energy_s(f) \times MOER_s
\end{equation}
where $energy_s(f)$ is the energy consumed by function $f$ and $MOER_s$ is the emissions per unit energy consumed.

\subsection{Data Center Location Selection}

Data centers are selected from a publicly available list of Google owned locations~\cite{googledata}. This choice is made in order to maintain feasibility in real world scenarios, since not every location is suitable for a data center.
Additionally, the locations we select span a geographically diverse range of the United States. This means data centers will have different balancing authorities, as well as a wide variety of solar energy profiles.

Table~\ref{tbl:my_label} summarizes the 9 locations we selected alongside the metrics associated with both their grid and grid-isolated energy profiles.

\subsection{Baseline Algorithms}

    \noindent
     \textit{OpenWhisk.} OpenWhisk is a modified variant of the consistent hashing algorithm and serves as a baseline for comparison to GreenWhisk. 

     \noindent
     \textit{Greedy Algorithm.} This algorithm uses a greedy approach to select servers with the minimum emissions or maximum carbon-free energy. We use this as a benchmark that considers solely carbon/energy awareness and not locality.

     \noindent
    \textit{Consistent Hashing.} This algorithm achieves locality-aware and load-aware balancing, even in situations where nodes are being added and removed from the system. It does so by mapping nodes and workloads to a point on a circle and comparing their distances. This algorithm provides a baseline performance model when OpenWhisk's algorithm is unavailable for experimentation, and it is well suited to intermittent power scenarios.

\section{Results}

\begin{figure}[t]
    \centering
    \includegraphics[width=\linewidth]{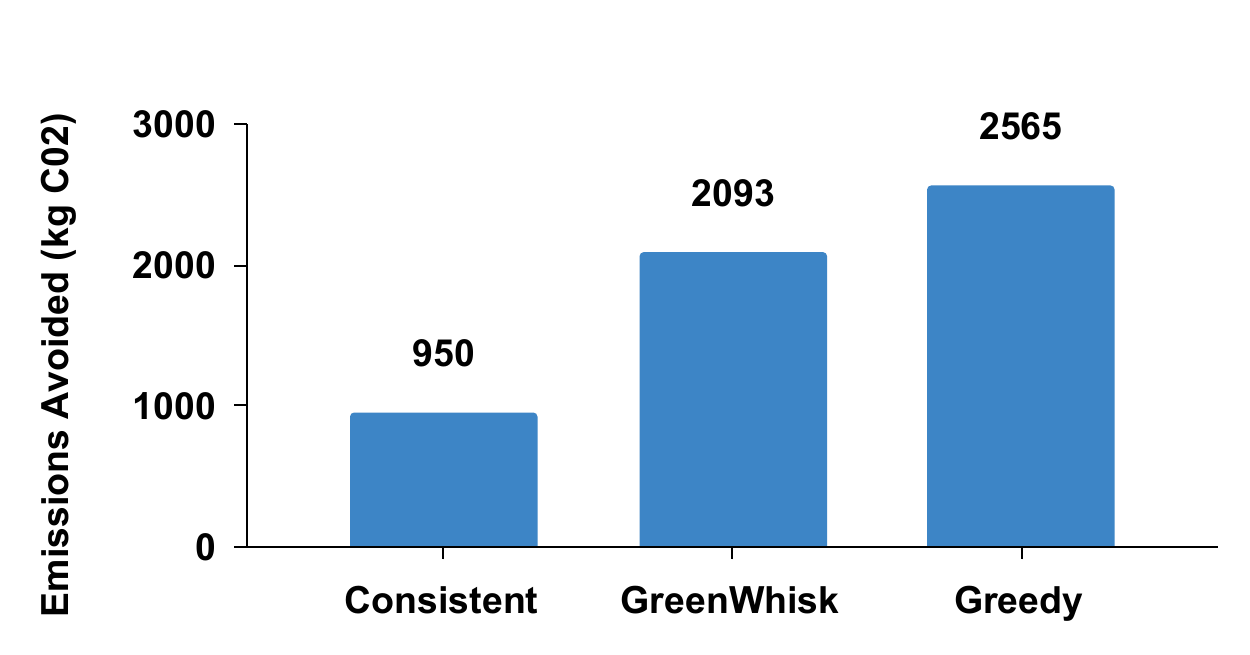}
    \caption{Emissions avoided by different baseline algorithms.}
    \label{graph:emissionsavoidedperalgorithm}
\end{figure}

\begin{figure}[t]
    \centering
    \includegraphics[width=0.99\linewidth]{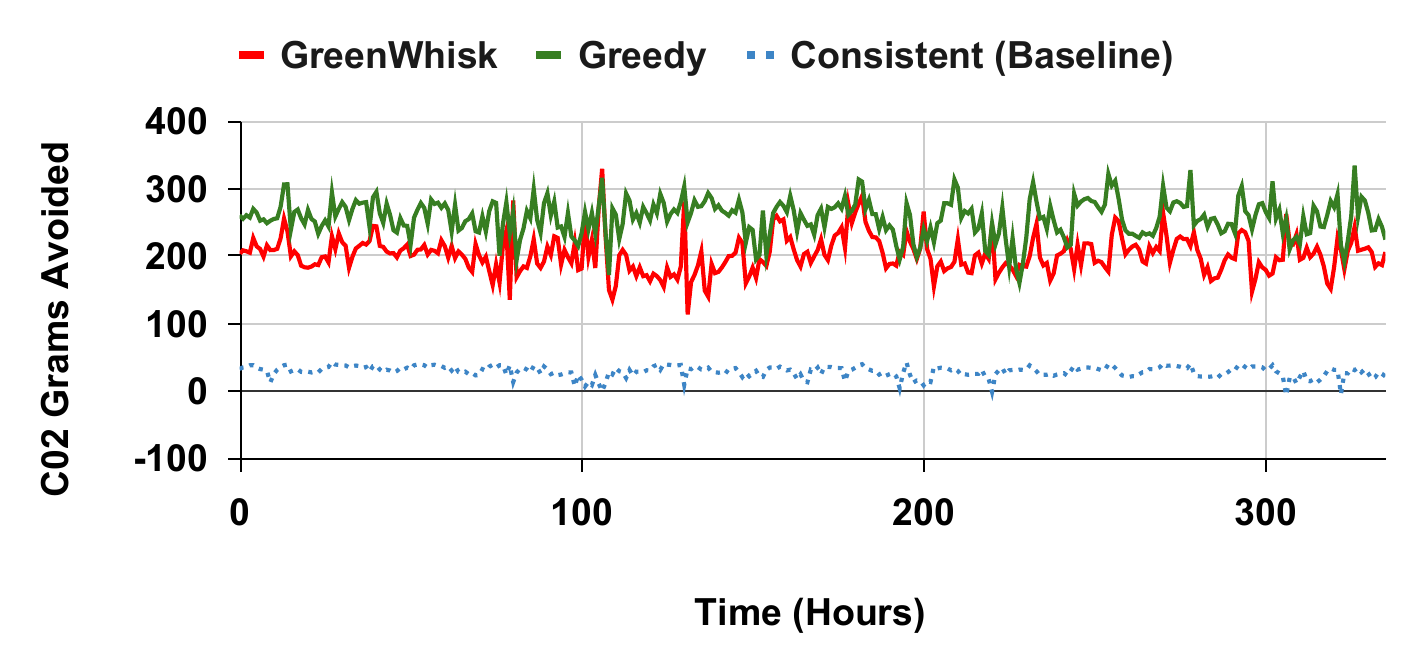}.
    \caption{Emissions avoided per hour for each algorithm.}
    \label{graph:emissionsavoidedday}
\end{figure}
\subsection{Grid-connected results}
We analyze the grid-connected mode, assuming servers are powered by the grid. 
\subsubsection{Emissions Avoided}
We evaluate the emissions avoided for different load balancing algorithms on a high workload dataset (Table~\ref{tbl:functrace}). Our simulation spans a year across 18 servers in 9 different locations, each hosting two servers. As our metric revolves around marginal emissions, all presented results are framed in the context of emissions avoided \textbf{over OpenWhisk}, signifying the emissions reduction achieved by relocating the workload to a different location. 
Figure \ref{graph:emissionsavoidedperalgorithm} shows the overall emissions avoided by different algorithms over the simulation period. We observe that all baseline algorithms performed better than OpenWhisk. The greedy algorithm exhibits higher emission avoidance as it prioritizes carbon reduction over locality. In contrast, consistent hashing primarily considers locality, resulting in lower emission reduction. GreenWhisk strikes a balance between both locality and emissions, providing a middle-ground solution.

Figure \ref{graph:emissionsavoidedday} illustrates how the emissions vary over time. The figure demonstrates that the consistent hashing algorithm may sometimes underperform compared to OpenWhisk. However, GreenWhisk and the greedy algorithm consistently outperform OpenWhisk, capitalizing on periods of low emissions.



\begin{figure}[t]
    \centering
    \includegraphics[width=\linewidth]{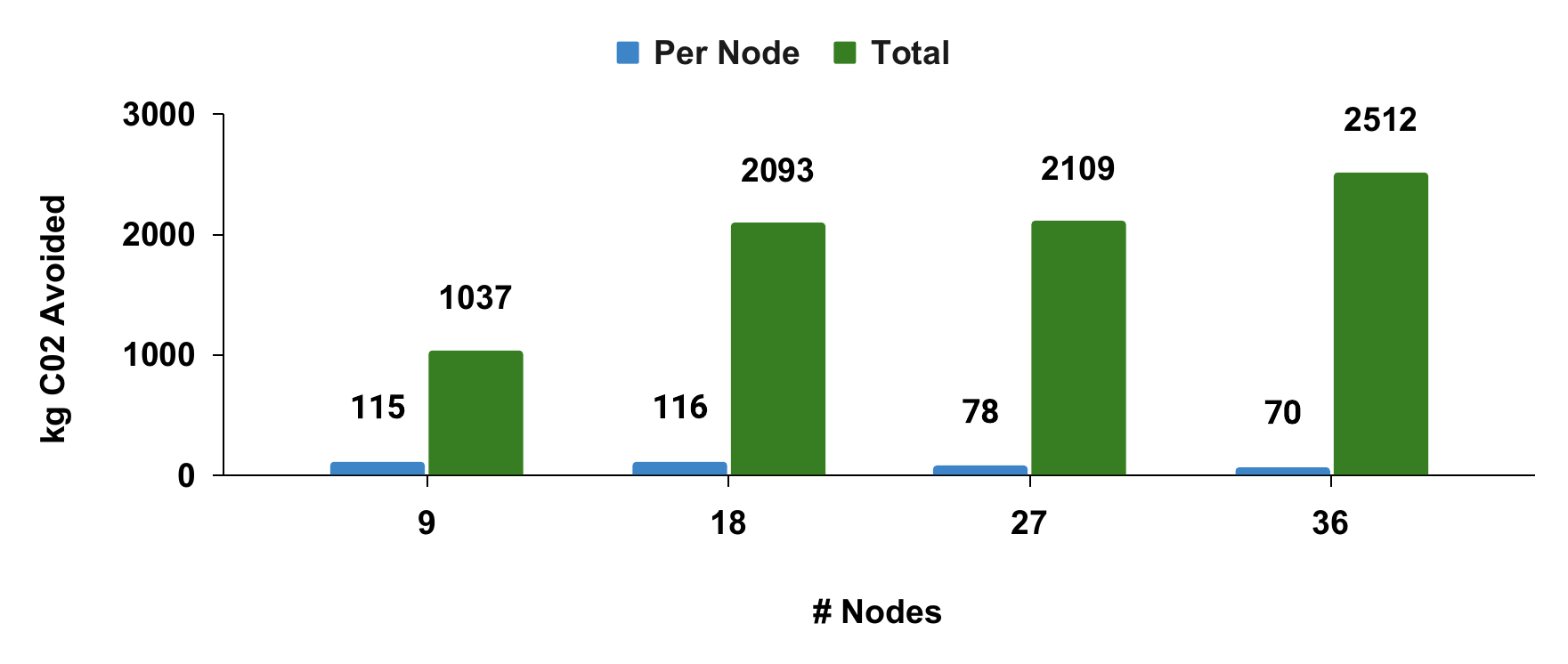}
    \caption{Emissions avoided for a varying number of server nodes. }
    \label{graph:emissionsavoidedpernodecount}
\end{figure}

\subsubsection{Impact of Servers}
We now analyze the impact on avoided emissions while varying the number of servers across nine locations, depicted in Figure \ref{graph:emissionsavoidedpernodecount}. Intuitively, as the number of servers increases, there is more opportunity for the algorithm to leverage spatial variations in carbon intensity. This trend is evident in the figure, where increasing the number of servers allows the algorithm to redirect workloads to low-carbon-intensity locations. As observed earlier, GreenWhisk outperforms when given more opportunities to switch a workload's target. Note that we use the same function trace for our simulations, implying that more servers become underutilized as we increase the node count. As such, it results in a decrease in avoided emissions per node. 


\subsection{Grid-isolated results}
We analyze the grid-isolated mode and assume servers are self-powered by solar and batteries. We assume each location has a 1kW solar array and a 3.8kWh battery shared by servers within each location. We sized solar and battery based on peak energy consumption and varied the solar and battery capacity if we increased or decreased the number of servers per location. 
\begin{figure}[t]
    \centering
    \includegraphics[width=3.1in]{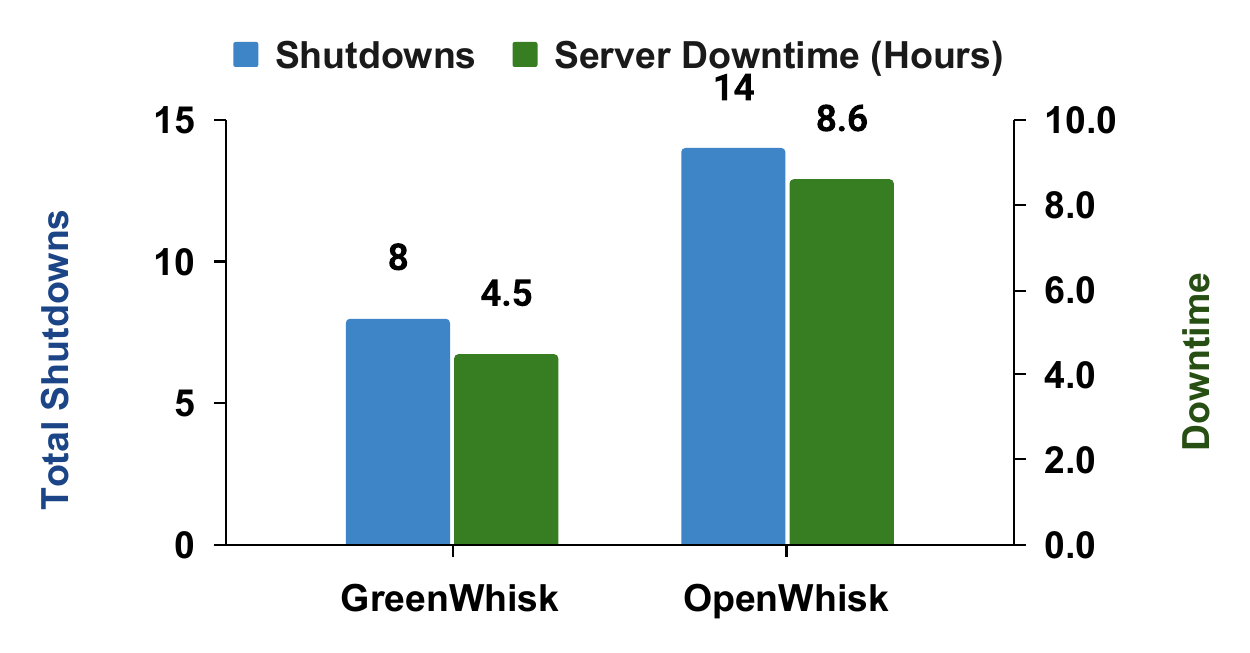}
    \caption{Availability comparison of solar-powered servers}
    \label{graph:solarstats}
\end{figure}
\subsubsection{System Availability}
Given that the servers are entirely powered by renewables, resulting in a zero carbon footprint, we shift our analysis to the availability of servers when running on intermittent energy. Figure~\ref{graph:solarstats} assesses server downtime and the frequency of shutdowns across these servers using GreenWhisk and OpenWhisk. Note that the downtime is aggregated per server and does not signify a scenario where {\it all} servers are offline.   Since GreenWhisk uses available energy as a metric to distribute workload, it ensures energy is uniformly consumed across all locations, thus reducing the number of shutdowns. OpenWhisk, on the other hand, uses locality to map the functions to the same server location, resulting in higher shutdowns.
In particular, we observe that GreenWhisk achieves a 50\% reduction in downtime and total shutdowns compared to OpenWhisk. This significant improvement highlights the effectiveness of GreenWhisk in minimizing server downtime and maintaining higher availability during intermittent energy conditions. 

\begin{figure}[t]
    \centering
    \includegraphics[width=2.8in]{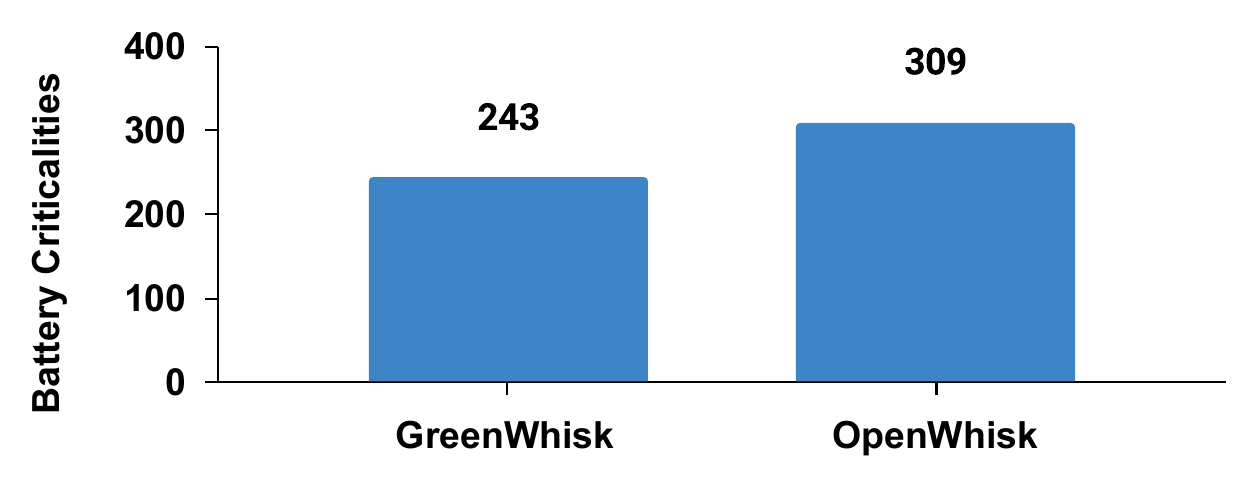}
    \caption{Comparison of critical battery levels}
    \label{graph:crit_states}
\end{figure}
\subsubsection{Battery Health}
Since operating batteries at low levels could degrade their performance over time, we also analyze how often the battery level falls below a certain threshold (20\% in our scenario). As illustrated in Figure \ref{graph:crit_states}, we observe that GreenWhisk experiences fewer instances where batteries go below the threshold. In other words, GreenWhisk maintains batteries at higher energy levels overall than OpenWhisk. This is because by using available energy to distribute workloads, GreenWhisk ensures even battery levels across servers, preventing the unnecessary depletion of a select few nodes when overall energy availability in the system is high.

\begin{figure}[t]
    \begin{subfigure}{\linewidth}
        \includegraphics[width=\linewidth]{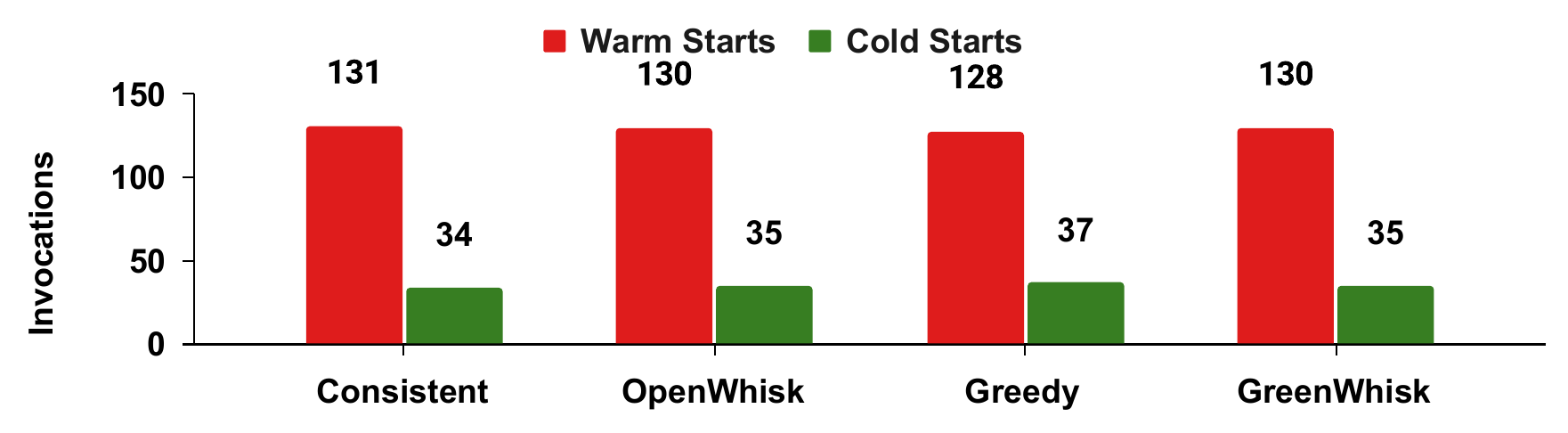}
        \caption{Rare dataset}  
        \label{graph:ric_invocationsub1}
    \end{subfigure}
    
    \begin{subfigure}{\linewidth}
        \includegraphics[width=\linewidth]{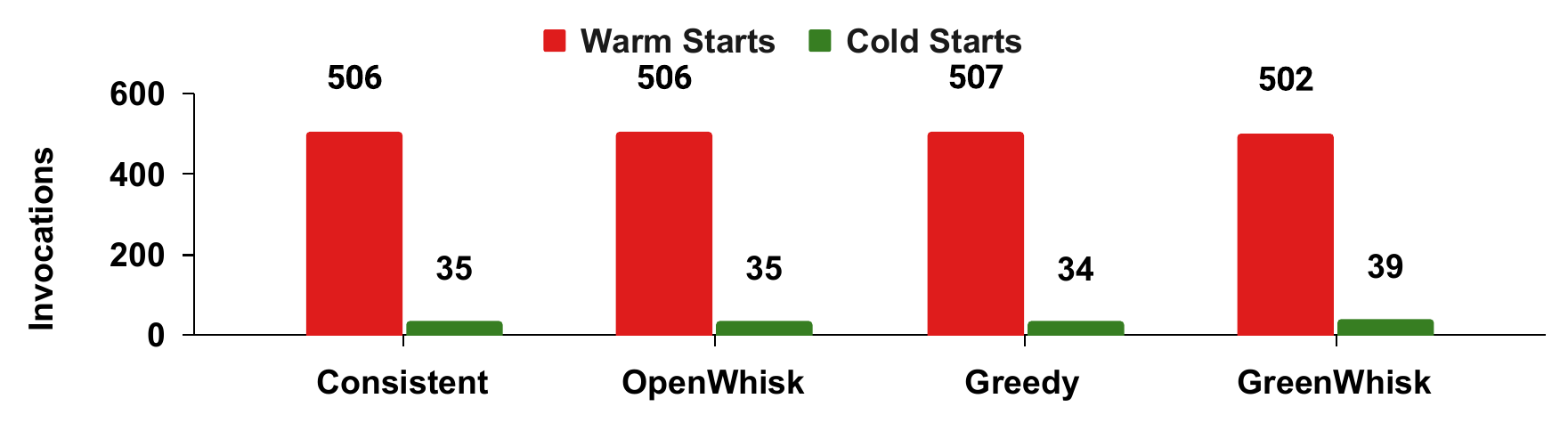}
        \caption{Medium dataset}
        \label{graph:ric_invocationsub2}
    \end{subfigure}

    \caption{Performance on Rare and Medium datasets on server cluster.}
    \label{graph:ric_invocations}
\end{figure}

\begin{figure}[t]
    \begin{subfigure}{\linewidth}
        \includegraphics[width=\linewidth]{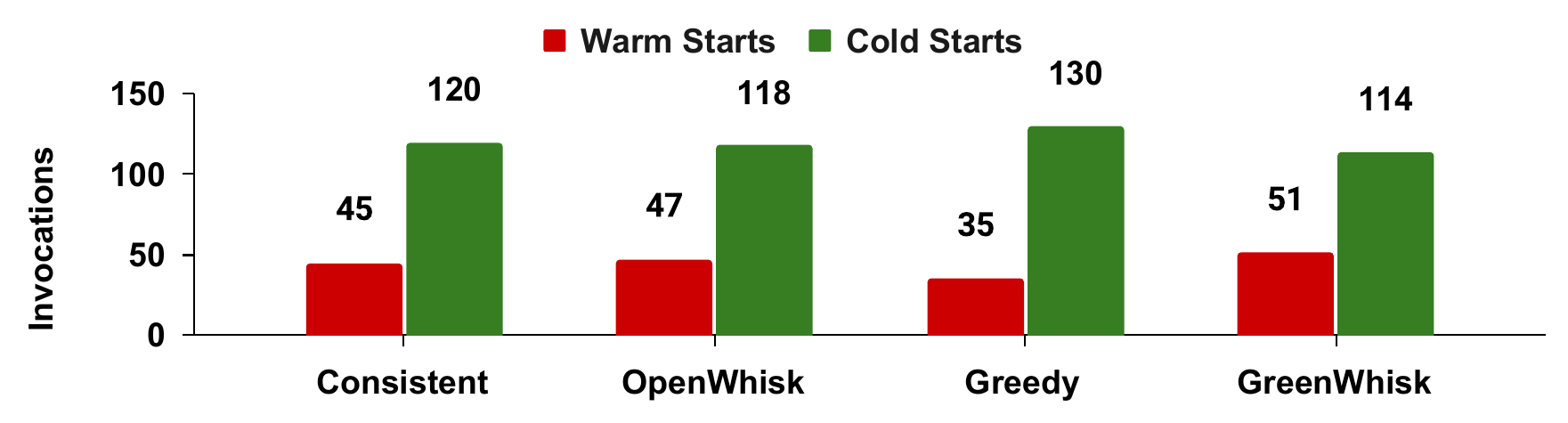}
        \caption{Rare dataset}  
        \label{graph:pi_rare}
    \end{subfigure}
    
    \begin{subfigure}{\linewidth}
        \includegraphics[width=\linewidth]{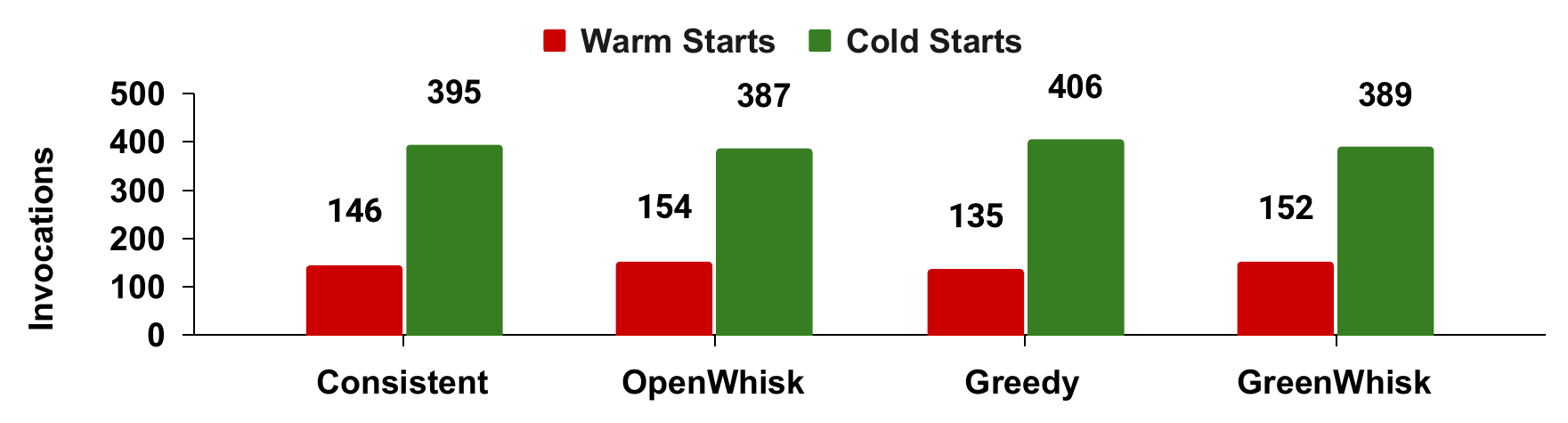}
        \caption{Medium dataset}
        \label{graph:pi_med}
    \end{subfigure}

    \caption{Performance on Pi cluster.}
    \label{graph:pi_invocations}
\end{figure}

\begin{figure}[t]
  \centering
    \begin{subfigure}{\linewidth}
        \includegraphics[width=3in]{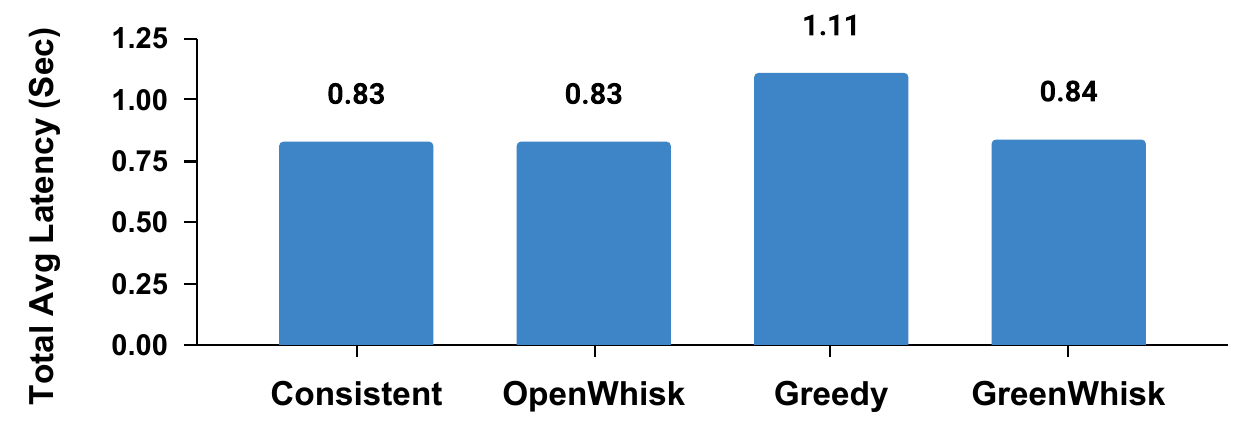}
        \caption{Server cluster}
\end{subfigure}
  \begin{subfigure}{\linewidth}
        \includegraphics[width=3in]{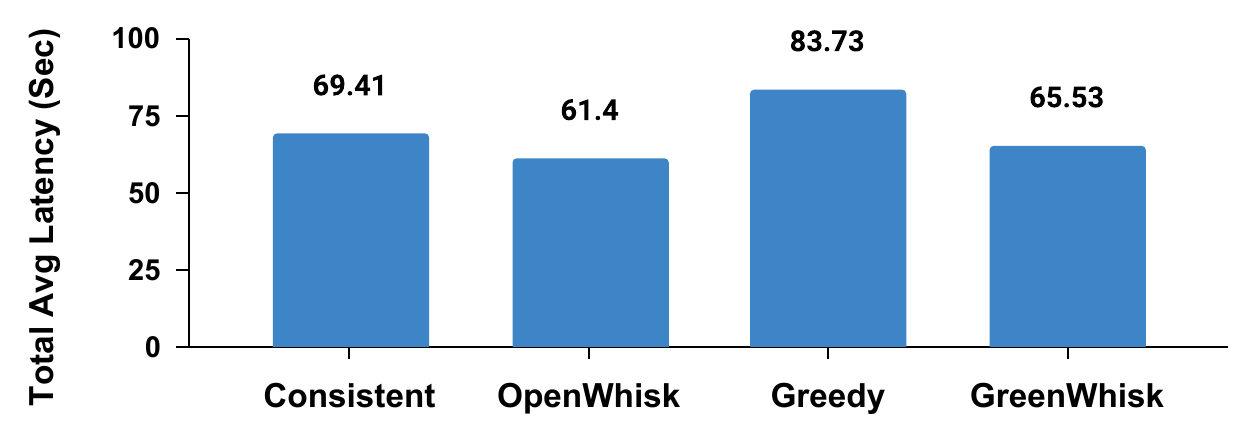}
        \caption{Raspberry Pi cluster}
\end{subfigure}
  \caption{End-to-end latency for different algorithms.}
    \label{graph:latency}
\end{figure}

\subsection{Emulation results}
We analyze the performance of various algorithms in our emulation setup. In our experiments, we execute the functions on server and Pi clusters and measure the overhead performance in executing these functions. 
\subsubsection{Performance}
Figures \ref{graph:ric_invocations} and \ref{graph:pi_invocations} compare GreenWhisk with other load balancing algorithms, focusing on cold start and warm start statistics across different datasets. The results indicate that GreenWhisk performs comparably to other baseline algorithms, demonstrating no significant degradation in performance despite its carbon awareness. As shown in Figure \ref{graph:ric_invocations}, the number of cold starts in the server cluster remains low across different datasets. However, the number of cold starts increases significantly when executed on the Pi cluster. This increase is likely due to limited memory, which prevents the system from running multiple concurrent containers. Nevertheless, compared to other algorithms, GreenWhisk still achieves fewer cold starts, indicating its effectiveness in maintaining locality awareness.

\subsubsection{Latency}
We also determine the end-to-end latency in function execution. Figure \ref{graph:latency} illustrates the average function execution latency for the algorithms on the medium dataset. Notably, GreenWhisk and OpenWhisk achieve the lowest latency, while the Greedy algorithm exhibits the highest latency. This latency difference can be attributed to the higher number of cold starts in the Greedy algorithm, as it lacks locality awareness. In contrast, OpenWhisk and GreenWhisk invoke more warm starts and experience fewer cold starts. Given that GreenWhisk is both locality and carbon-aware, the impact on latency is not substantial. We also observe higher latency when switching from server clusters to Pi clusters. Since the Pi clusters are not as powerful, much of the increased latency can be attributed to executing the functions on these less powerful devices. Overall, we observe that the GreenWhisk's latency is similar to other baseline algorithms. 

\begin{figure}[t]
    \includegraphics[width=3.1in]{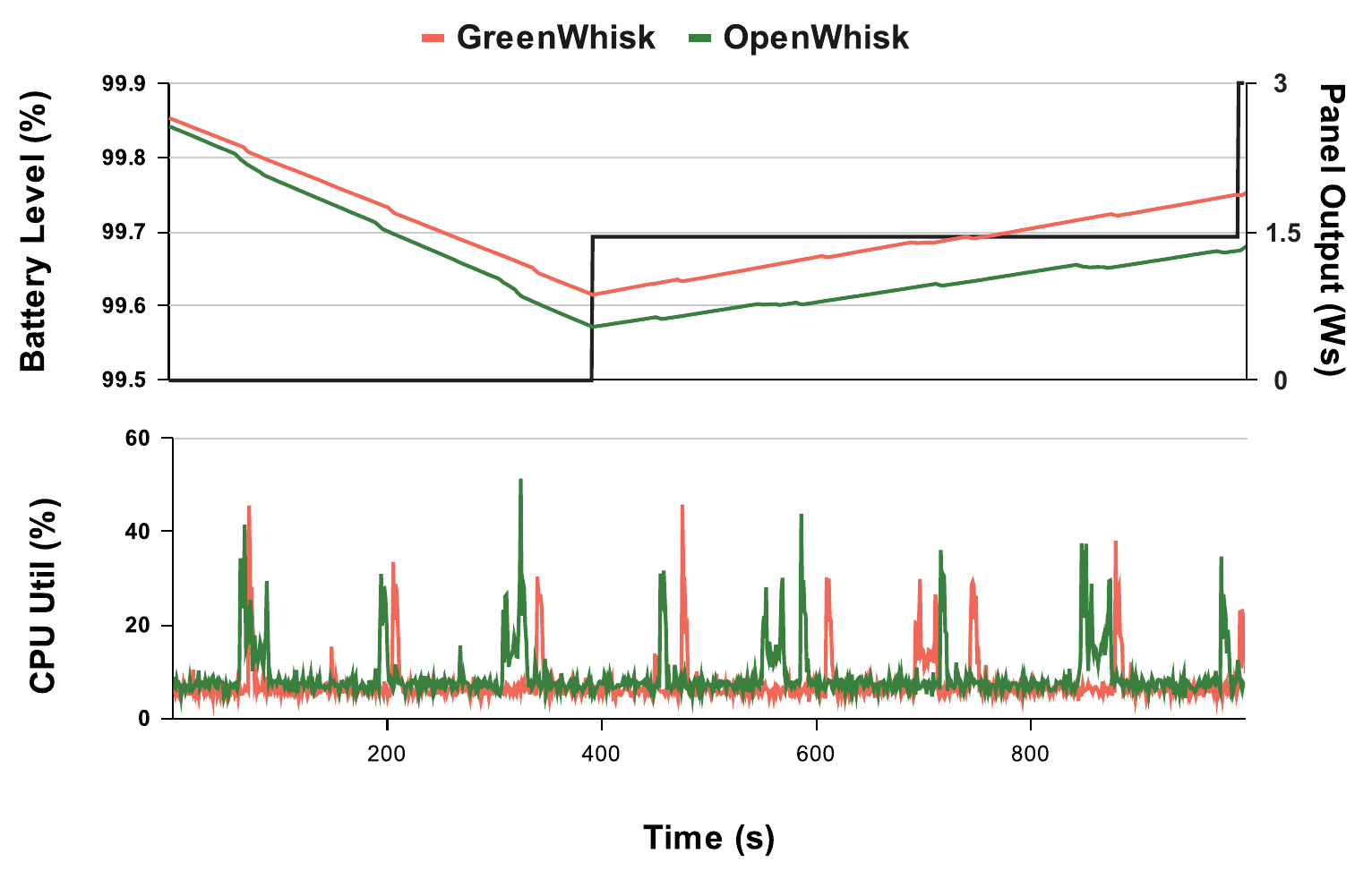}
    \caption{Energy and utilization on one of the invokers.}
    \label{graph:utilization}
\end{figure}

\subsubsection{Energy and Utilization}
Figure \ref{graph:utilization} illustrates the energy and utilization patterns of a server, including solar output, battery level, and CPU utilization, collected from our emulation on the Pi cluster. Since GreenWhisk accounts for available energy in the invoker, the figure shows lower battery depletion within the invoker. In other words, GreenWhisk assigns less workload to this invoker compared to OpenWhisk, which uses locality to map functions to the same server. We also observe that the algorithm charges the battery when excess energy is available. Notably, around the 400 second mark, an increase in the overall battery level coincides with the availability of solar energy. We also analyzed the available battery energy across different servers (not shown in the figure) and observed that GreenWhisk depletes energy evenly, ensuring a more balanced distribution of workload and energy consumption across the cluster.

\section{Related Work}

{\bf Serverless Computing.}
Serverless computing is gaining traction across various application scenarios, including machine learning training \cite{liu2022funcpipe, gimeno2022mlless, jiang2021towards, wang2019distributed,cassel2022serverless}, and distributed computation\cite{jonas2017occupy}. There has been significant work in optimizing serverless platforms for these scenarios, including model partitioning \cite{yu2021gillis}, batch processing \cite{ali2020batch}, cold start optimization \cite{yang2022infless}, and various other methodologies \cite{sampe2018serverless, fuerst2022locality}.
In contrast, our work focuses on designing a carbon-efficient serverless platform, and we leverage grid and energy characteristics to reduce carbon footprint.

\noindent
{\bf Energy Management. } 
There have been numerous studies on reducing the energy consumption of cloud platforms and data centers~\cite{liu2011greening,palasamudram2012using,chen2005managing}. Prior works have also explored enhancing energy efficiency by enabling applications to control their energy and carbon footprint \cite{shen2013power, liu2008chameleon,  souza2023ecovisor}. For instance, applications can constrain power usage per container based on their workload \cite{shen2013power}. Recently, there has been work on virtualizing power~\cite{deng2012policy, wang2014case} and providing virtualized energy component control to allow applications~\cite{souza2023ecovisor}. In contrast, our approach does not involve explicit power control. Instead, we expose this information to the platform and introduce mechanisms for high-level algorithms to optimize carbon efficiency.

\noindent
{\bf Carbon Management. }
Researchers have studied using renewable energy and batteries to power computing systems~\cite{palasamudram2012using, govindan2011benefits, goiri2013parasol,agarwal2021redesigning, rezaei2023solar}. 
Research studies leveraging batteries typically emphasize reducing energy costs~\cite{govindan2011benefits, palasamudram2012using}. 
Moreover,  existing work on renewable integration and batteries has been conducted in the context of distributed storage~\cite{sharma2011blink}, and Hadoop~\cite{goiri2012greenhadoop}. This process involves degrading an application's quality of service to reduce energy consumption. Similarly, several efforts have been made to optimize carbon efficiency that harnesses both temporal and spatial variations in carbon intensity~\cite{bashir2021enabling, gupta2016cool, lee2015cost, hanafy2023carbonscaler, wiesner2021let}. In essence, these techniques involve strategically scheduling workloads to capitalize on low carbon intensity periods. 
With the increase in the adoption of renewable energy, fluctuations in carbon intensity are expected to increase in both grid-connected and grid-isolated scenarios~\cite{radovanovic2022carbon, xing2023carbon}. Our work enables handling these fluctuations within the context of a serverless platform.

\section{Conclusion}
Addressing carbon intermittency in software systems has gained significant importance as part of broader efforts to mitigate the environmental impact of computing. Particularly for systems relying on renewable energy sources, transparently managing the intermittency inherent in renewables becomes crucial.  In this context, our work introduces GreenWhisk, a carbon-aware serverless computing platform designed to effectively handle the challenges posed by carbon intermittency in both grid-connected and grid-isolated scenarios. 
Our mechanisms allow for the creation and integration of various carbon-aware load balancing algorithms. Our extensive emulation and simulation results demonstrate that the performance overhead introduced by our mechanisms is minimal. Furthermore, our carbon-aware load balancing algorithms show that it can reduce carbon emissions compared to other baseline algorithms. 

{\bf Acknowledgements.} We thank IC2E reviewers for their valuable comments, and WattTime for access to their carbon-intensity data. This work is supported in part by NSF grants \#2324873, Pitt Momentum Funds, and Mascaro Center for Sustainable Innovation.


\bibliographystyle{plain}
\bibliography{paper}

\end{document}